# A Link-based Approach to Entity Resolution in Social Networks

Gergo Barta[1]

[1]Department of Telecommunications and Media Informatics, Budapest University of Technology and Economics, Magyar tudosok krt. 2. H-1117 Budapest, Hungary
barta@tmit.bme.hu

## Abstract

*Social networks initially had been places for people to contact each other, find friends or new acquaintances. As such they ever proved interesting for machine aided analysis. Recent developments, however, pivoted social networks to being among the main fields of information exchange, opinion expression and debate. As a result there is growing interest in both analyzing and integrating social network services. In this environment efficient information retrieval is hindered by the vast amount and varying quality of the user-generated content. Guiding users to relevant information is a valuable service and also a difficult task, where a crucial part of the process is accurately resolving duplicate entities to real-world ones. In this paper we propose a novel approach that utilizes the principles of link mining to successfully extend the methodology of entity resolution to multitype problems. The proposed method is presented using an illustrative social network-based real-world example and validated by comprehensive evaluation of the results.*

## 1. Introduction and Related Work

Undoubtedly Facebook is the most popular and extensively used social network currently. While Google+ is gaining ground and others like LinkedIn, Orkut and Badoo also have their respective market share, these solutions tend to specialize their services on a narrow group of users instead of taking on Facebook eye to eye.

Pages on Facebook are edited and maintained by users only, there is no central administration. Some editors are closely related to the topic of the page like the owner of a service or manufacturer of a product, other individuals are mere enthusiastic fans of the given topic. There is no objective way to tell these pages apart, nor there is to rank or rate the value and utility of pages based on their content. Due to the social philosophy of Facebook there is absolutely no restriction to creating duplicates of the same topic, any supporter of a specific subject can create his own page in the topic and start discussion and information sharing. The result is a vast number of duplicate pages with a little extra flavor added to each variant.

In this environment any analysis has to be preceded by troublesome data cleaning to work efficiently and accurately. Entity resolution is the process of deduplicating data references that refer to the same underlying real-world entities. It is particularly important when performing data cleaning or when integrating data from multiple sources. Besides corporate master data management for customer and product information an obvious example is price engines, e.g. Shopzilla or PriceGrabber. These services offer easy selling price comparison of a single product offered by multiple vendors. A real-world example of duplicated records is shown in Table 1. While this action is relatively easy for humans and is carried out on-the-fly, machines however, without the proper understanding of information encapsulated in each record, tend to stumble in resolving the issue. The high importance and difficulty of the entity resolution problem has triggered a large amount of research on different variants of the problem and a fair amount of approaches have been proposed, including predictive algorithms [1], similarity functions [2], graphical approaches [3] and even crowdsourcing methods [4].

Table 1. A real-world price engine example

| ID | Product Name | Price |
|----|--------------|-------|
| 1 | Apple iPad 4 (16 GB) Tablet | $589.00 |
| 2 | iPad 2 16 GB with WiFi in | $313.45 |
| 3 | Apple iPad Air WiFi 32GB | $379.99 |
| 4 | Sealed Apple Ipad4 Wi-fi 16gb | $379.99 |
| 5 | Apple iPad 2, 3, 4 Aluminum | $31.95 |
| 6 | Apple iPad 64 GB with WiFi | $897.52 |
| 7 | Apple iPad Air (64 GB) Tablet | $857.87 |
| 8 | Apple iPad 2 16 GB with WiFi | $488.99 |

Two main varieties were presented in the past by researchers of the subject: similarity-based and learner-based methods.

Similarity-based techniques require two inputs, a similarity function *S* that calculates entity distance and also a threshold *T* to be applied. The similarity function $S(e_1,e_2)$ takes a pair of entities as its input, and outputs a similarity value. The more similar the two entities are according to the function, the higher the output value is. The basic approach is to calculate the similarity of all pairs of records. If $S(e_1,e_2) \geq T$ is true for the given *{e₁,e₂}* entity pair, they are considered to refer to the same entity. Gravano et al. show in [5] that cosine similarity and other similarity functions are efficiently applicable where data is mainly textual.

Learner-based techniques handle entity resolution as a type of classification problem and build a model *M* to solve it. They represent a pair of entities *{e₁,e₂}* as a *[a₁,...,aₙ]* vector of attributes, where $a_i$ is a similarity value of the entities calculated on a single attribute. This approach, as any learner-based method, requires a training dataset *Tr* in order to build *M*. A suitable training set contains both positive and negative feature vectors corresponding to matching pairs and non-matching pairs respectively. The classifier built this way can later be applied to label new record pairs as matching or non-matching entities. Formally the entity pair *{e₁,e₂}* resolve to the same real-world entity if *M(Tr, e₁,e₂, [T/CM])* delivers true. In this case providing threshold *T* is optional, as it can be calculated based on a cost matrix *CM* as well. Sehgal et al. [6] successfully implement learners like logistic regression and support vector machines to enhance geospatial data integration. While Breese et al. [1] used decision trees and Bayesian networks for collaborative filtering. From now on $e_1 \leftrightarrow e_2$ denotes both $S(e_1,e_2)$ and *M(Tr, e₁,e₂, [T/CM])*.

The rest of the paper is organized as follows: in Section 2 an intriguing application is presented for motivational reasons then problem formulation and experimental setup is discussed. In Sections 5 and 6 we go into detail on data preparation and ER methods to be applied, while results are presented in Section 7. Finally, we conclude in Section 8.

## 2. A MOTIVATING EXAMPLE

We will now motivate the problem of entity resolution in a specific domain, social networks, and highlight some of the issues that surface using an illustrative example from the telecasting domain. Consider the problem of trying to construct a database of television programs broadcasted on major channels and their respective "fanpage", a place for sharing information and discussion, collected from Facebook, one of the most popular social networks.

Table 2. Example of an EPG record (a) and its respective fanpage (b)

(a)

| Id | Day | Title | Category | Start | Stop | Channel | Subtitle | Description |
|---|---|---|---|---|---|---|---|---|
| 605 | 21-10-2013 | Britain from Above | Nature | 13:00 | 14:00 | BBC HD | null | Documentary series in which broadcaster Andrew Marr… |

(b)

| Id | Name | Link | Category | Likes | Website | Talking about |
|---|---|---|---|---|---|---|
| 109…813 | Britain From Above | www.facebook.com/pages/Britain-From-Above/109…813 | Tv show | 282 | www.bbc.co.uk/britainfromabove | 3 |

While Facebook enlists an extensive amount of extremely useful and informative pages, the lack of central regulation and multitude of duplicate topics make information retrieval cumbersome. Finding a way to isolate the most suitable page for each television title is not only a practical service for users, but also an algorithmically difficult task. Apart from the actual content, fanpages have very few descriptive attributes, and the same goes for television programs acquired from an EPG[1]. Some attributes of a television show extracted from an EPG and its respective fanpage to be recommended are shown in Table 2, note some marginal variables are omitted. Performing ER in such an information poor environment is a tolling task.

For any given television show there is a number of suitable fanpages. A fanpage is considered suitable if its content is closely related to the show. The definition of "the most suitable" page is strictly subjective, although there are several ground rules to be considered during fanpage selection.

- *Rule 1*: Under no circumstances should an unrelated page be recommended.

- *Rule 2*: When more than one page is eligible, the most suitable is selected.

Violating *Rule 1* would lead to user distrust, while the violation of *Rule 2* degrades the quality of service. In context to *Rule 2* the most suitable fanpage is the one most interwoven with the subject and also with the most intensive discussion going on.

---

[1] Electronic Program Guide

Entity resolution in general is a delicate, human input intensive error prone process. In contrast to our example generic ER methods work on a single entity type only. As we have seen in Section 1 generic methods use attribute similarity functions to calculate entity-level similarity. Since there are two types of entities discussed here (television program and fanpage) attribute matching is not so straightforward. Human input is crucial to identify corresponding attributes, and find a way to establish *hooks* between the rest, a way of connecting entities. This knowledge is also very application-specific, there is no proper way to automatically establish the *hooks*.

## 3. PROBLEM FORMULATION

As opposed to generic ER this example enlists two distinct entity types; type A and B corresponds to television programs and social network fanpages respectively. Consider the graph example *G* displayed in Figure 1 where the two distinct entity types form the vertices. To identify related vertices edges (*E*) are added to the graph, this information is provided by the blocking method (see Section 4 for more). Optionally the edges can also be weighted based on the confidence level of the similarity function chosen. The *G(A,B,E)* graph is bipartite, meaning there are no edges between identical types. A↔A similarity (comparing TV titles) is out of the scope of this paper, while B↔B similarity (between fanpages) is only observed as a function of their ties with their respective type A entity. For example if both $A_3$↔$B_7$ and $A_3$↔$B_8$ delivers a positive answer then due to the transitivity of the ↔ operator also $B_7$↔$B_8$ resolves to the same entity weighted by the two edges involved. This approach, called link mining, has been researched extensively (refer to the survey by Getoor et al. [7]). The achievements of link mining and link-based ranking are used in algorithms like PageRank and HITS, and are consequently utilized in the way we select the suitable candidates in the post-processing step.

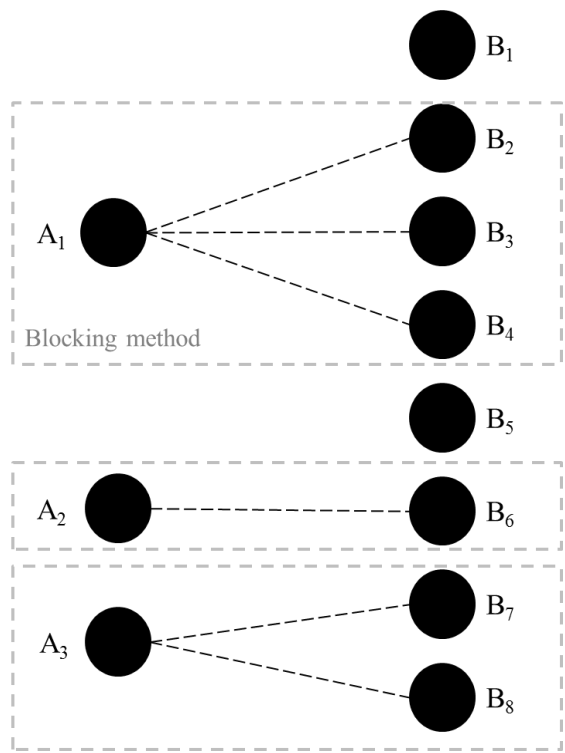

Figure 1. The telecasting example as a bipartite graph

## 4. EXPERIMENTAL SETUP

Our experiments were based on data gathered by XMLTV[2], a publicly available backend to download television listings. We opted for selecting the schedule of four major channels (BBC News, BBC HD, Al Jazeera English and Animal Planet), and the time period was set to October 2013. The query returned 1110 records to be used in our experiments, due to the innate program structure of television channels, there is substantial amount of repetition both on a daily and weekly basis. Since the Facebook API takes only a single query expression, the title of the television program, to avoid unnecessary querying and processing caused by recurring titles, all electronic program guide records are aggregated on *{Channel, Title, Duration}* level. It is safe to assume that the *{Channel, Title, Duration}* tuple is a unique identifier for all television shows. All other query parameters are being set independently of the respective television program.

Aggregating television programs to avoid unnecessary queries results in substantial compression of the experimental database, 163 unique television shows are extracted of the original 1110 titles. As of 2014 there are 54 million fanpages accessible on Facebook[3], comparing all titles with all available pages would be utterly unfeasible. To avoid the computational explosion caused by comparing all available candidates, generic ER approaches use a blocking method, a heuristic to identify the most probable candidates in order to cut search space significantly. In our case querying the Facebook search engine is a convenient way of blocking, as it delivers the most probable pages based on the internal context of Facebook (see Fig. 1).

Executing subsequent Facebook queries for every unique show title presents us with a total 258 fanpage results. That gives us the number of record pair comparisons to make after applying the blocking method. Figure 2 shows the number of results returned and their respective frequency. Keep in mind that the number of results per query is limited to 10 as part of the blocking mechanism. Although the high number of queries with a single result is promising, we should, by no means, automatically accept and recommend those results as it could violate *Rule 1* presented in Section 2. In any case, evaluation of query results is required.

---

[2] See http://sourceforge.net/projects/xmltv/
[3] See http://www.statisticbrain.com/facebook-statistics

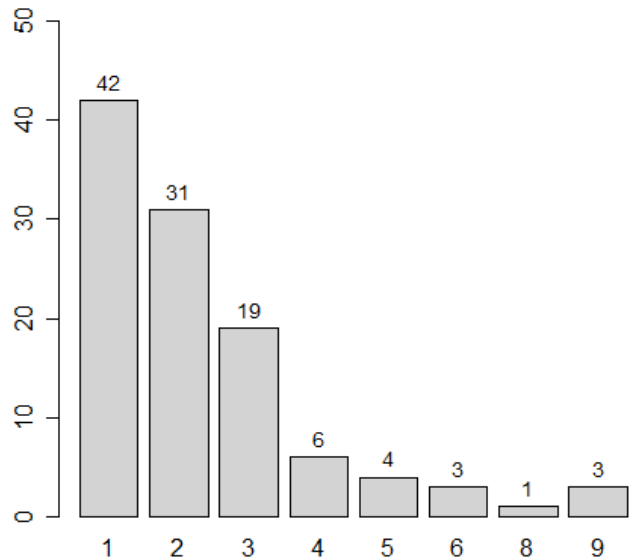

Figure 2. Distribution of result frequency

## 5. ENTITY SIMILARITY MEASURES

As we have seen in Section 2 in order to successfully identify matching entities of distinct types certain connections or *hooks* are to be set up. This step requires substantial human input. Also Table 2 shows that very few attributes are available to do so. In case of string attributes (*Title* and *Name*) edit distance and related string functions can be used. Long-winded strings containing whole paragraphs (e.g. *Description*) are to be processed to isolate a set of proper nouns and compare those only. Polynomial variables like *Category* need preprocessing as well, since they are from different sources distinct variable values are to be matched first.

As we have seen before the low number of entity attributes account for an information poor environment. Utilizing any additional data source can boost the performance of the ER process greatly. Since the whole process is run on-line, as the resolving method relies heavily on Facebook, other on-line services can be included as well (e.g. Google Search to find the URL of each channel). With the help of these additional services new variables can be computed.

As the variable *Link* is available, incidentally the main output of the whole process, the content of the fanpage can be queried (the same goes for *Website*). Page content then can be extensively used to generate *hooks*, in this case these are flags (yes-no questions) regarding the actual page content and especially the descriptive About section (since that is publicly available). New variables include whether the *Channel* variable or its URL is explicitly mentioned on the page, if the website link referenced by the fanpage is the actual website of the channel, whether the linked website is referencing the *Channel* and its URL and *Title* at all.

There are a total of 12 new descriptive features created. Flags have a value of either 0 or 1, while distance like features such as *Title* edit distance have continuous values on the [0,1] interval. Numeric variables are normalized in the same interval as well.

# 6. RESOLVING ALTERNATIVES

In Section 1 we discussed two well-known options for solving ER problems. In our work we explored the possibility of extending both of them to 2-type ER problems (see Figure 3).

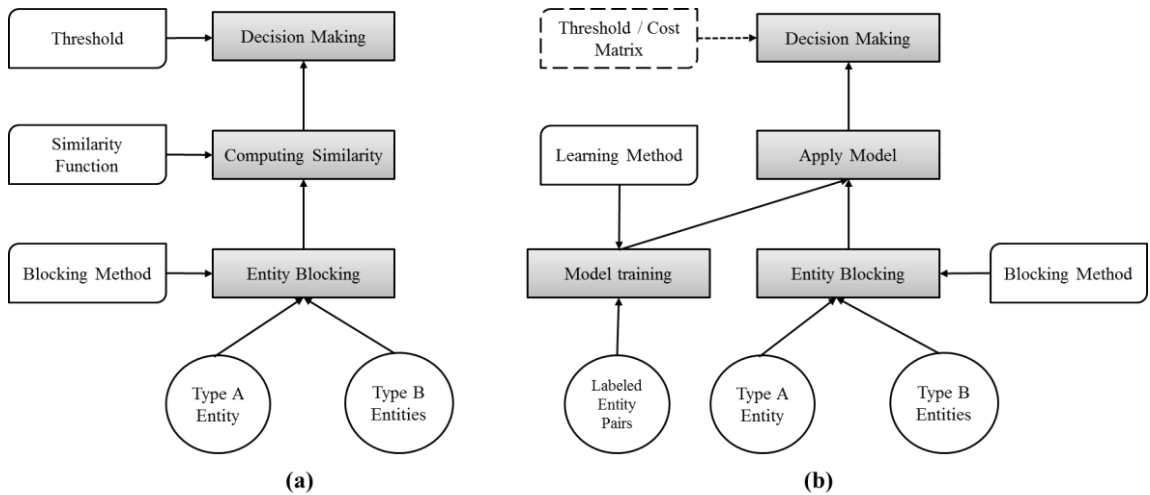

Figure 3. Similarity-based (a) and learner-based (b) techniques extended to multi-type ER

We opted for one of the most common implementation for the similarity-based method by using a simple score model. This is quite straightforward since all our features are already transformed into the same interval. On some group of features a maximum was computed first in order to avoid bias as they referenced the same underlying attributes only in slightly different aspects. Otherwise by a simple addition the final score is calculated (on a 10-point scale) and a threshold can be applied. For this setup threshold was manually set to 5 points, which covered the majority of the examples (see Figure 4 for details).

The learner-based approach treats ER as classification problem using the feature vector generated in the previous section. We explored several different classifier models implemented in RapidMiner (RM) a free data mining toolset[4]. Out of the many possibilities 3 were selected based on their performance and penetration; Logistic Regression (LR), Support Vector Machine (SVM) and Random Forest (RF). Both the LR and SVM implementation in question was done by Stephen Rueping called myKLR and mySVM respectively (see [8]) RM also offers automatic threshold adjustment based on model confidence, costs and ROC analysis.

---

[4] See http://www.rapidminer.com

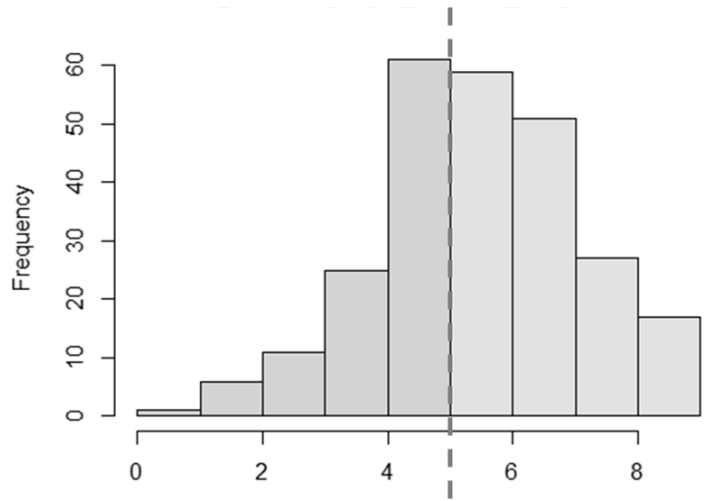

Figure 4. Similarity-based approach: respective score frequency and threshold application.

## 7. RESULTS AND DISCUSSION

Our experiments were carried out on the telecasting dataset using the four different decision making method described in the previous section. A dataset was prepared by labelling entries manually for training and testing purposes. For the sake of result comparability we elected three common known performance measures in machine learning; recall, precision and F-measure. In case of learner-based methods training was done using 10-fold cross-validation with stratified sampling and a cost matrix was provided with a 20% extra penalty for the so-called type I misclassifications (false positive examples).

As seen in Figure 5a all approaches delivered comparable results when looking for "the most suitable" fanpage. Learner-based methods performed very similarly, with RF falling behind in precision only. The score model dominated in Recall suggesting a relatively low threshold, but F-measure compared fairly well to learning methods.

Tests were also carried out after relaxing *Rule 2*; so that all related fanpages would be acceptable. In Figure 5b all methods but RF show a substantial drop in recall caused by the foreseeable increase of false negatives. Compensated by increased precision F-measure increased for all classification models as well, yielding an almost identical performance. The more rigid approach of the similarity-based scoring method however failed to satisfy the needs of the modified environment.

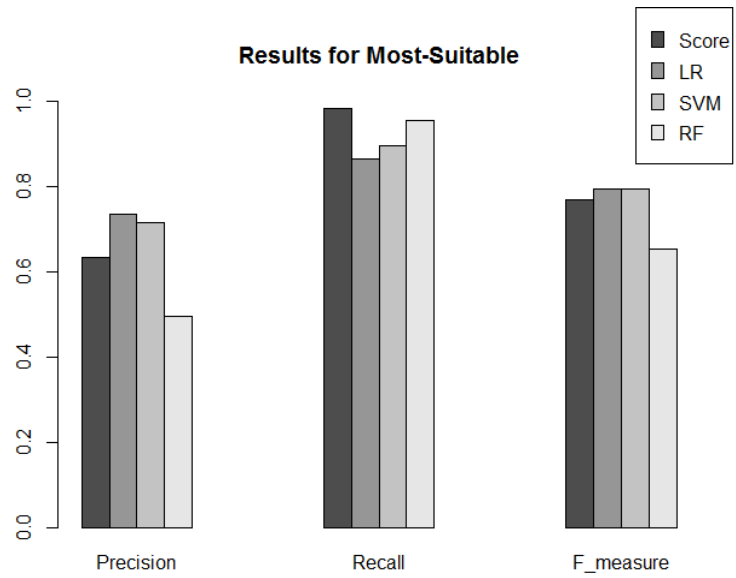

(a)

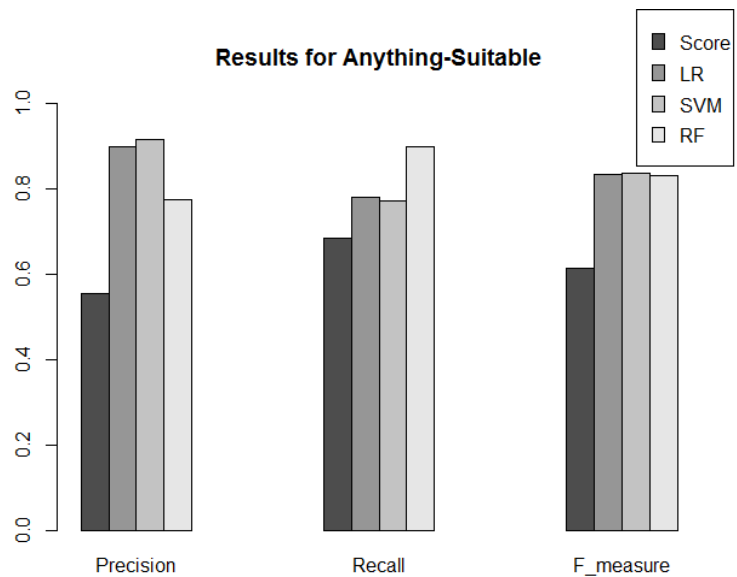

(b)

Figure 5. Comparing the results of the "most suitable" (a) and "anything suitable" (b) methods

## 8. CONCLUSIONS

In this paper we proposed a method to successfully extend entity resolution to multitype problems through an interesting application domain. This particular solution eases information retrieval in social networks, an increasingly growing field for analysis. The procedure utilizes the achievements of link mining to effectively bound entities of distinct types. We performed an extensive evaluation of the proposed methods on an illustrative real-world problem and conclude that the promising results can lead to growing scientific interest and may contribute to valuable services in the future.

**Authors**

Gergo Barta received the MSc degree in computer science in 2012 from the Budapest University of Technology and Economics, Hungary, where he is currently working toward the PhD degree in the Department of Telecommunications and Media Informatics. His research interests include spatial data analysis, data mining in interdisciplinary fields, and machine learning algorithms.

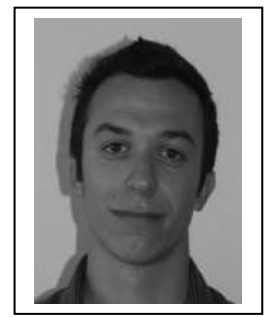